# Notes from the Physics Teaching Lab: A Magneto-Mechanical Harmonic Oscillator

Kenneth G. Libbrecht[1]

Department of Physics, California Institute of Technology

**Abstract.** We describe a magnetically driven torsional oscillator that is well suited for teaching the physics of simple harmonic motion using a collection of hands-on, quantitative experiments. The mechanical Q of the system can be tuned using eddy-current damping, while optical read-outs provide electronic signals than can be recorded using nothing more than a basic digital oscilloscope. The 40-Hz oscillator is described by simple harmonic motion to high accuracy, providing many satisfying comparisons between theory and experiment.

## Introduction

We designed and built the Magneto-Mechanical Harmonic Oscillator (MMHO) for teaching the physics of the Simple Harmonic Oscillator (SHO) at an introductory level, with a particular emphasis on high-Q oscillators and clocks. Because harmonic oscillators play such a central role in modern physics, and high-Q mechanical clocks (in the form of micro-machined quartz crystal oscillators) are omnipresent in modern electronic devices, the MMHO experiment introduces students to a fascinating and vital area of physics and technology.

While this topic has been explored in physics teaching laboratories many times over the years (with [1995Jon, 2017Car, 2018Kha, 2020Mar, 2026Lel] being some recent examples), the MMHO focuses on what can be done with a precision instrument using a magnetic drive, optical readouts, and other features that connect well with modern digital electronics. By exploring fundamental physics concepts using basic electronics test equipment that can be found in many research labs, the MMHO provides a practical hands-on experience for students who are interested in careers in experimental physics and technology.

Beyond introducing laboratory techniques and concepts that mesh well with theory courses, we have also strived to make the MMHO instrument easy and fun to use, with an open and intuitive feel. The hardware is ruggedly built, and the oscillator parameters were chosen with student interaction in mind. Using the MMHO apparatus, students can collect accurate data quickly and reliably, while learning about many aspects of SHO physics through a variety of quantitative and qualitative investigations. This document describes and illustrates the hardware in the MMHO and describes numerous experiments that can be performed with it. Although the MMHO is not commercially available, our hope is that this document may prove beneficial for teaching-lab instructors looking to incorporate an SHO instrument in their curriculum.

---

[1] *kgl@caltech.edu*

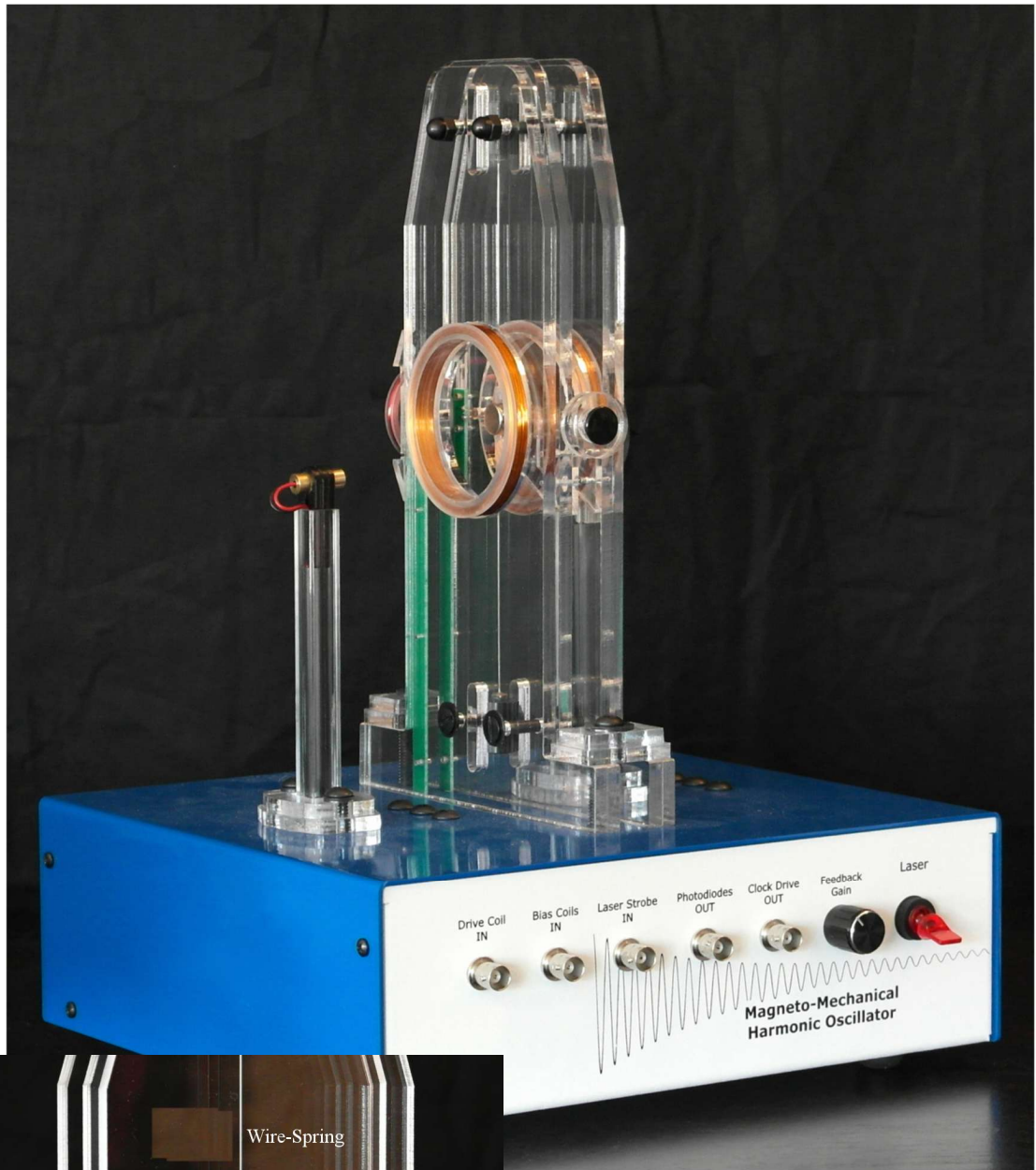

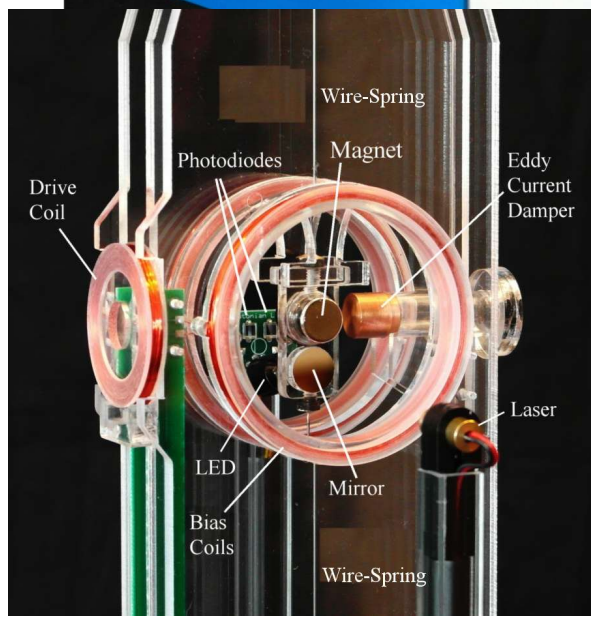

*Figure 1. (Above) The Magneto-Mechanical Harmonic Oscillator (MMHO) consists of an acrylic "tower" (29 cm high) sitting atop an electronics chassis with a footprint of 26x28 cm. Additional acrylic posts on either side of the tower hold a diode laser module (seen in this image) and a photodiode assembly mount (behind the tower in this photo).*

*(Left) This annotated image shows a close-up of the essential elements of the torsional oscillator, described in detail in the text. Appendix 1 discusses additional hardware specifications.*

# Instrument description

Figure 1 shows the overall structure of the MMHO instrument, along with a closer view of its main operating elements. At the heart of the instrument is a torsional oscillator that consists of a 12.7-mm-diameter cylindrical rare-earth magnet supported by two vertical steel wires. The wires both support the test-mass magnet assembly and allow it to rotate about the vertical axis, while twisting of the wires provides a restoring torque. The primary torsional mode behaves very much like an ideal harmonic oscillator with a natural resonance frequency of about 40 Hz. The removable eddy-current damper can adjust the mechanical $Q$ of this oscillator from about 100 to 3000.

The torsional oscillation mode can be excited using the Drive Coil, which is typically used to produce an oscillating magnetic field at a frequency around 40 Hz. At the position of the test-mass magnet, the Drive-Coil magnetic field is mainly horizontal and perpendicular to the magnetic moment of the permanent magnet (that points along its cylindrical axis). The applied torque on the test mass is then

$$\tau_{drive}(t) \approx \vec{\mu} \times \vec{B} \approx \mu B(t) \tag{1}$$

where $\vec{\mu}$ is the magnetic moment of the test mass and $\vec{B}$ is the applied magnetic field. In the small-angle approximation applied here, the drive torque is essentially independent of the angular position of the test mass about the vertical axis.

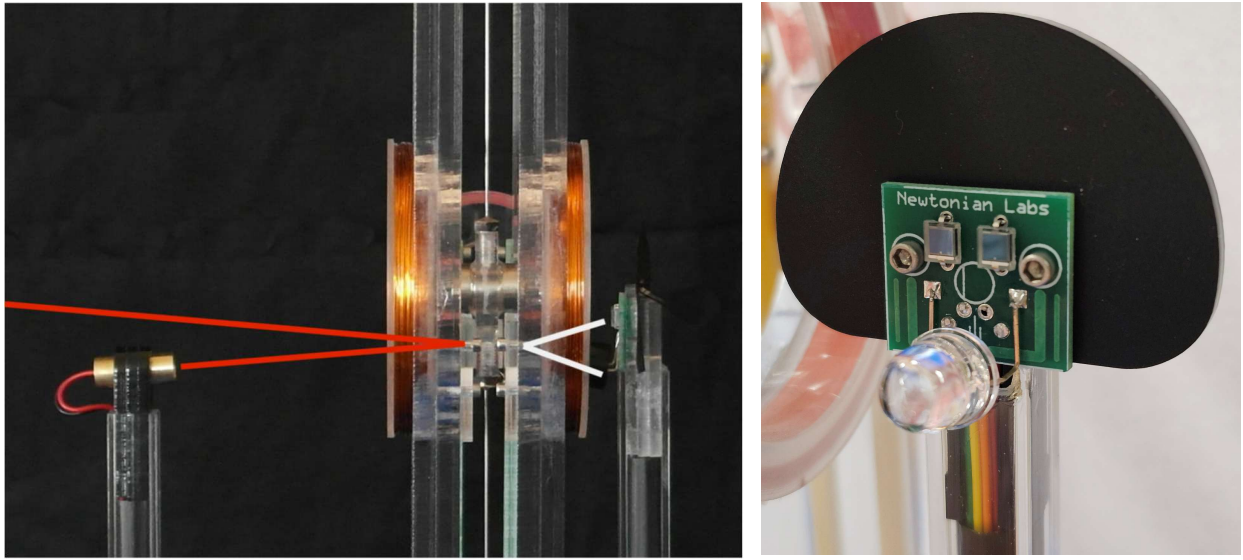

*Figure 2. (Above Left) This side view of the MMHO tower illustrates a laser beam from a diode-laser module (red line added to photo) reflecting from one of the test-mass mirrors. When the test mass is oscillating at 40 Hz, this beam makes a horizontal streak of light on a plastic ruler, giving a visual indicator of the oscillation amplitude. On the other side of the tower, light from an LED reflects from the other test-mass mirror onto a photodiode pair (white line added to photo).*

*(Above Right) This small circuit board holds a large LED and a pair of small rectangular photodiodes. The beam of light from the LED (roughly collimated by the lens-like plastic housing) reflects from the second test-mass mirror (left photo) and then onto the photodiode pair. In the small-angle limit, the difference of the two photodiode signals is proportional to the test mass rotation angle about the vertical axis. The rounded black plastic part blocks the oscillating LED light beam from shining into people's eyes.*

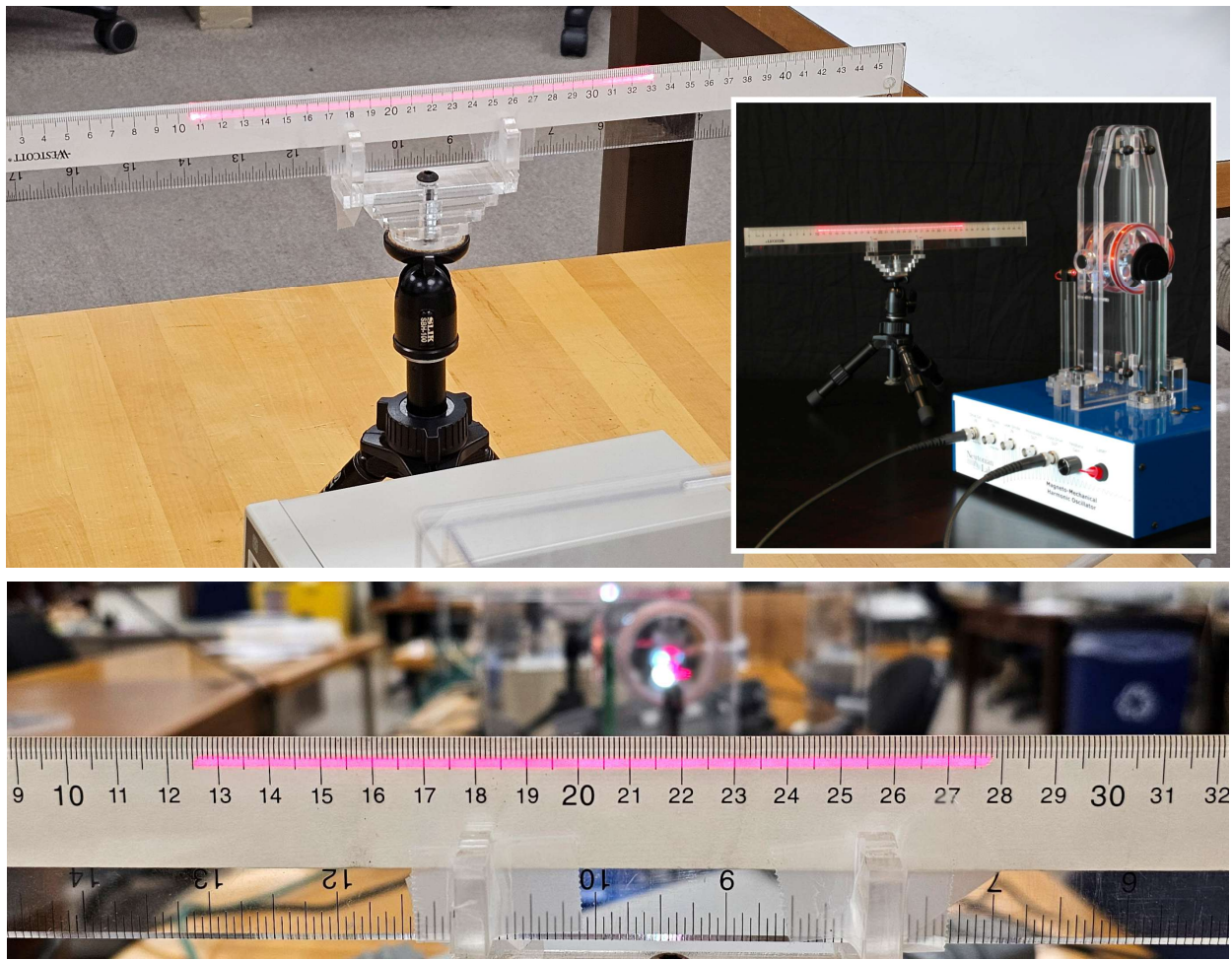

*Figure 3. These photos show how the laser streak can be used to measure the amplitude of the torsional oscillator. At 40 Hz, the laser looks like a streak of red light to the naked eye, and the length of the streak can be measured to an accuracy of 1mm using nothing more than a plastic ruler. The laser streak also provides students with a handy at-a-glance indicator of the oscillator status. It is also possible to strobe the laser light using a basic function generator, which yields a shorter streak of light that moves back and forth at a frequency equal to the difference between the strobe frequency and the oscillator frequency.*

**The laser streak.** The diode-laser module seen in Figures 1 and 2 provides an appealing visual indication of the oscillation amplitude of the test mass, with views like those shown in Figure 3. The red laser beam reflects off one of the small mirrors below the test mass magnet, so the oscillating mirror produces a streak of red light on a clear plastic ruler that has some white paper tape attached. The bright streak of light is useful as a status indicator for the state of the oscillator, plus it provides an obvious method for measuring the oscillator's amplitude using the ruler. Converting the light streak to radians requires nothing more than simple geometry, and students can easily estimate the accuracy of the $\tan\theta \approx \theta$ small-angle approximation in this system (where $\theta$ is the angular position of the MMHO torsional oscillator).

**The photodiode signal.** Figure 2 also shows an LED and photodiode-pair that can be used to measure the oscillation amplitude using an electronic signal. In a nutshell, the LED produces a diffuse but roughly Gaussian "beam" of light that is roughly collimated by the LED's lens-like plastic housing. The LED beam reflects off a second mirror on the test-mass assembly and onto a matched pair of photodiodes on a small printed-circuit board. The two photocurrents are subtracted in the MMHO electronics (see Appendix 1), and we call the resulting difference signal the "photodiode signal" or just the PD signal.

When the oscillator amplitude is small, the PD signal is essentially a simple sinusoidal voltage with the amplitude of the sinusoid being proportional to the amplitude of the MMHO test mass oscillations. The linearity breaks down at high amplitudes, however, and Figure 4 shows how the PD waveform becomes distorted in the process. In practice we avoid these nonlinear PD signals when making quantitative observations, but explaining them is a good teaching moment.

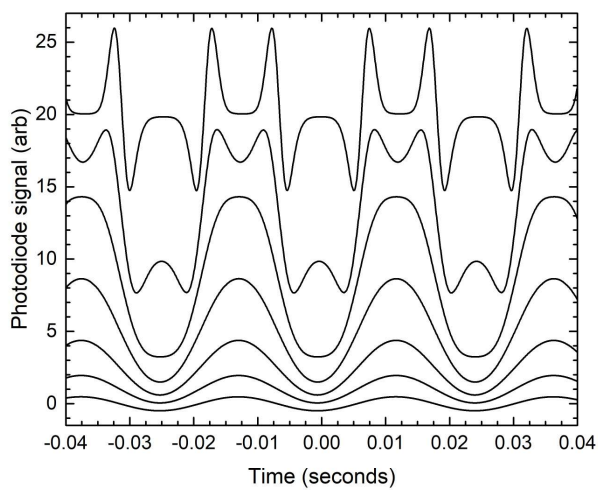

Figure 4. These waveforms demonstrate how the photodiode signal becomes distorted at high MMHO amplitudes. In all cases the signals average to zero over time, so the different traces were vertically displaced here for clarity. The curves range from low oscillator amplitudes at the bottom of the graph to high amplitudes at the top.

The shapes of the different PD signals can be qualitatively understood from the geometry of the photodiode setup in Figure 2, and making this connection is a good exercise for students. In practice, however, we avoid these distorted signals and limit quantitative observations to low amplitudes, where the sinusoidal signal amplitude is proportional to the MMHO amplitude.

Figure 5 shows how the MMHO oscillation amplitude varies with the Drive Coil current (near 40 Hz), as measured using both the laser streak and the PD signal. At any drive frequency, the MMHO amplitude is expected to depend linearly on the Drive Coil current, and this linear trend is clearly seen in the data over nearly four orders of magnitude. Except for the highest few points in Figure 5, the MMHO can be well described by the small-angle approximation.

## MMHO experiments

We designed the MMHO to exhibit an oscillator behavior that is as close as possible to that of a perfect one-dimensional damped harmonic oscillator. The basic SHO is a staple of physics education, so it is nice when students can see that behavior reproduced with high fidelity in the teaching lab. The MMHO equation of motion is simply that for a torsional harmonic oscillator, which we write as

$$I\ddot{\theta} + \gamma\dot{\theta} + \kappa\theta = \tau_{drive}(t) \qquad (2)$$

where $\theta(t)$ is the angle of the test-mass magnet relative to its position at rest, $I$ is the mass moment of inertial of the test-mass assembly, $\kappa$ is the restoring torque provided by the wire supports, $\gamma$ is a damping constant, and $\tau_{drive}(t)$ is provided by the Drive Coil (see Equation (1)).

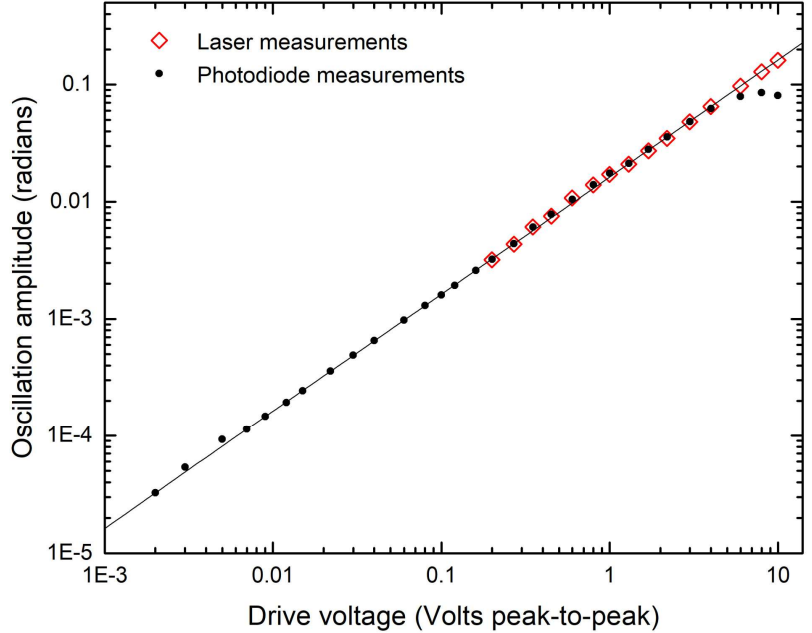

Figure 5. These data compare the MMHO oscillation amplitude measured using the laser streak (red points) and the PD signal (black points). The length of the laser streak was converted to radians using basic geometry, and the overlap between the two data sets was used to calibrate the PD signal.

Note that the PD signal amplitude depends linearly on oscillator amplitude except at the highest amplitudes, where the PD signals look like the examples shown in Figure 4.

In the absence of any damping or external torque, the resonant oscillation frequency of the oscillator is simply

$$\omega_0 = \sqrt{\frac{\kappa}{I}} \tag{3}$$

Including damping, we prefer to work with the dimensionless quality factor $Q$, in this case defined by

$$Q = \frac{I\omega_0}{\gamma} \tag{4}$$

Solving the equation of motion for a freely decaying oscillator (with no applied torque) gives a ringdown behavior with

$$\theta(t) = \theta_0 e^{-t/T_{ringdown}} \cos(\omega_d t + \phi) \tag{5}$$

where

$$T_{ringdown} = \frac{2Q}{\omega_0} \tag{6}$$

and

$$\omega_d = \omega_0 \sqrt{1 - \frac{1}{4Q^2}} \tag{7}$$

The constants $\theta_0$ and $\phi$ are dictated by the initial conditions at the start of the ringdown. Because highly damped oscillators are not so much fun to work with, we designed the MMHO to operate with $Q > 100$, which means that $\omega_d = \omega_0$ to an accuracy of 12ppm, which is smaller than we can observe.

In the discussion that follows, we will mostly switch to frequency units with $f = \omega/2\pi$, because making measurements in cycles/second is the norm in experimental physics.

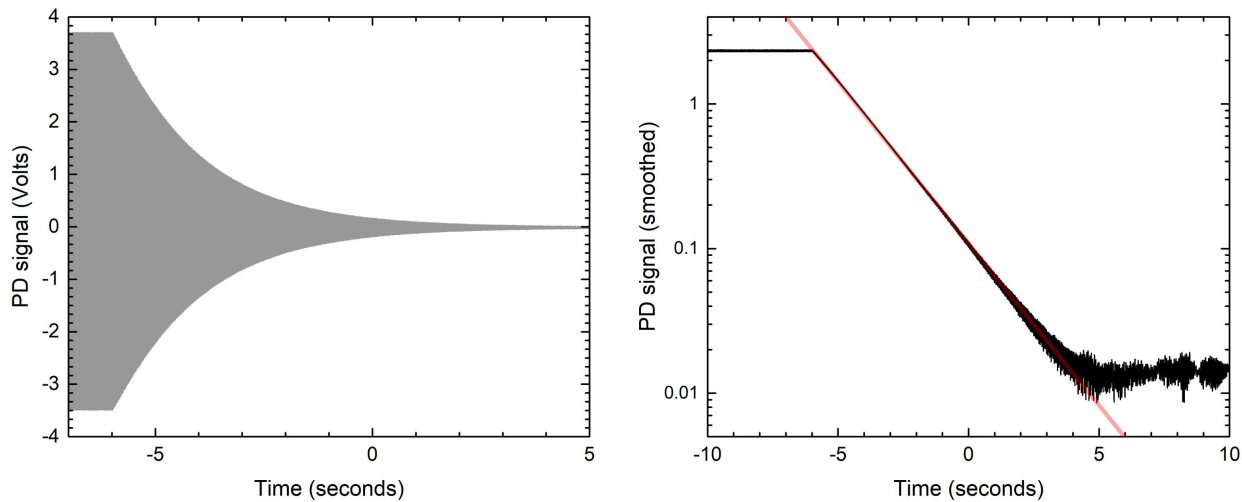

*Figure 6. These data demonstrate the MMHO ringdown observed on an oscilloscope operating in Roll mode. (Left) This graph shows the raw data pulled from the 'scope, starting with the oscillator driven at a steady-state amplitude. About one second in this graph, the drive was turned off while the oscilloscope captured the PD signal as a function of time. With 10,000 points recorded, individual oscillations can be seen the raw data but are not resolved in this plot. (Right) This plot shows the same PD data analyzed to produce a Q measurement. We took the absolute value of the raw PD data and then smoothed it in post processing to generate $\langle |\theta| \rangle (t)$. The red line shows a fit exponential with $T_{ringdown} = 1.94$ seconds, from which we obtain $Q = 247$. The value of Q can be adjusted by changing the position of the eddy-current damper in the MMHO (see Figure 1).*

## Ringdown measurements

Several different experimental methods can be used to measure the ringdown behavior of the MMHO, with varying degrees of difficulty and accuracy. With a high-Q oscillator like the MMHO, a direct ringdown measurement is usually the best way to determine Q.

### Ringdown on the oscilloscope

With a modern digital oscilloscope, the MMHO ringdown is most easily observed by putting the 'scope in roll mode and simply recording the PD signal, as demonstrated in Figure 6. This measurement technique is simple and intuitive, although the subsequent analysis requires handling a lot of data points and using some kind of smoothing algorithm. Using a 12-bit oscilloscope generally improves the dynamic range of the measurement (compared to an 8-bit 'scope).

### Ringdown using data-logging DMM

The oscilloscope method becomes somewhat laborious when Q is high and the ringdown time is long. In this case, the Q is best measured using a data-logging digital multimeter (DMM) like the Keysight 34461A or the Siglent SDM3045X. Although this method requires additional hardware, it also teaches students about these useful measurement tools, and the data collected require little further analysis. One simply records the AC amplitude of the PD signal as a function of time and fits the data to an exponential decay, as demonstrated in Figure 7.

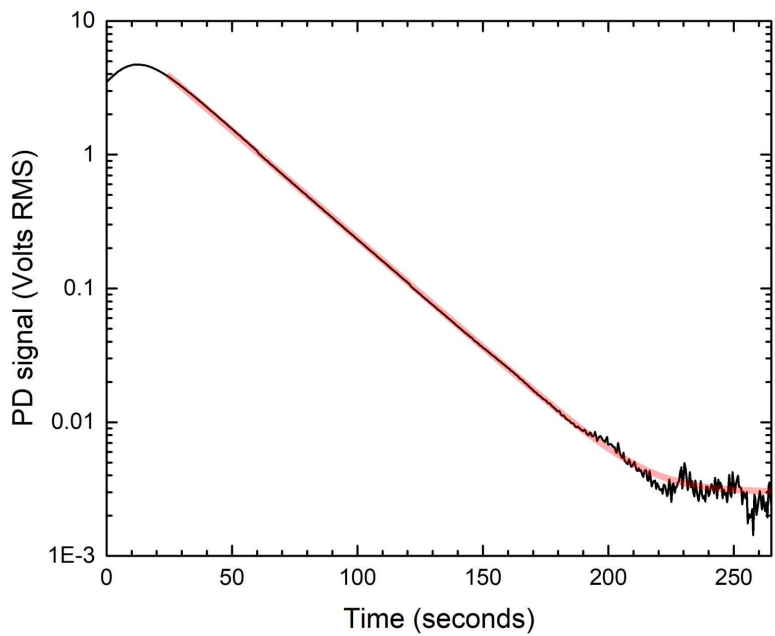

*Figure 7. To produce this plot, we simply recorded the AC amplitude of the PD signal as a function of time using a data-logging digital multimeter. In this case the eddy-current damping was removed, so the ringdown time is much longer than in Figure 6. This plot also started with the MMHO amplitude quite high, so the PD signal exhibited nonlinear behavior at the earliest times in the plot. The red line shows a fit exponential with $T_{ringdown} = 26.8$ seconds and $Q = 3421$. To approximate the background level in the theory curve, we included a 3mVrms constant signal added in quadrature to the main exponential.*

### Ringdown measured from the laser streak

Just to show that it can be done, Figure 8 demonstrates a ringdown measurement using only measurements of the length of the laser streak as a function of time. This method is surprisingly accurate, but its educational content is clearly not as high as the other methods.

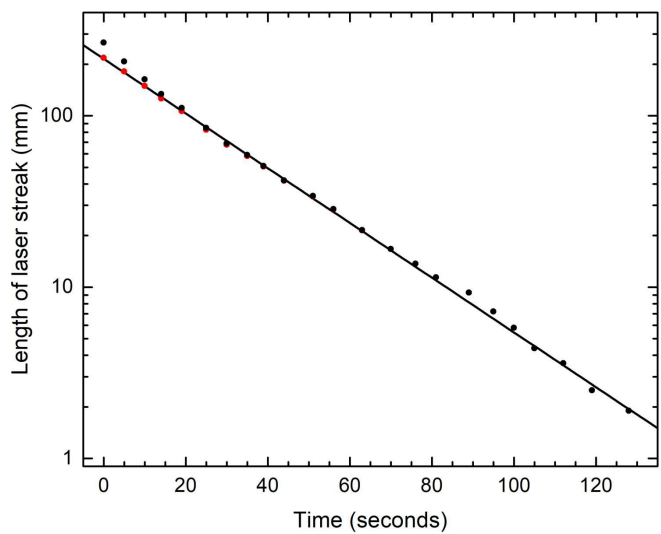

*Figure 8. This plot shows the length of the laser streak on a plastic ruler (see Figure 3), extracted from still images taken from a video of a MMHO ringdown. The black points show the direct measurements and the red points include a geometrical large-angle correction. This measurement method is somewhat laborious and not as educational as the other methods, but it can produce quite accurate results if one is careful.*

## Driven harmonic oscillator response

Another staple of SHO physics is to apply a sinusoidal drive force to an oscillator and examine its steady-state response [2020Rus, 2018Kha]. For the MMHO, we used an external function generator to produce an applied torque $\tau_{drive}(t) = \tau_0 \cos(2\pi f_0 t)$, and solving the equation of motion gives the steady-state solution $\theta(t) = \theta_0(f) \cos(2\pi f_0 t + \varphi(f))$, where

$$\theta_0(f) = \frac{\tau_0 Q}{\kappa} \frac{f_0^2/Q}{\sqrt{(f^2 - f_0^2)^2 + \left(\frac{ff_0}{Q}\right)^2}} \tag{8}$$

In this expression the second quotient equals unity when $f = f_0$, so we see that the maximum oscillator amplitude on resonance is proportional to $Q$.

When $Q$ is high, the above expression near resonance can be approximated by

$$\theta_0(f) \approx \frac{\tau_0 Q}{\kappa} \frac{f_0/Q}{\sqrt{4(f - f_0)^2 + \left(\frac{f_0}{Q}\right)^2}} \tag{9}$$

In this limit, the oscillator energy takes the form of a Lorentzian function of frequency (see by squaring the above function). and the oscillator phase is

$$\varphi(f) \approx \frac{\pi}{2} + \tan^{-1}\left(\frac{2Q(f - f_0)}{f_0}\right) \tag{10}$$

Note that the phase goes to zero when $f \ll f_0$ (so the oscillator response is in-phase with the applied torque) and goes to 180 degrees when $f \gg f_0$ (so the oscillator is 180 degrees out-of-phase with the applied torque). As with the ringdown, we can measure the MMHO driven response using several different methods.

### From the laser streak

Visually estimating the length of the laser streak on a ruler as a function of the drive frequency yields surprisingly good results, as illustrated in Figure 9. With the eddy-current damper in place, it takes a few seconds for the MMHO to settle to its steady-state amplitude after each frequency change, and this is conveniently about the same time it takes to record a data point. This is a quick and useful exercise for beginning students, who are often surprised by the accuracy of this method. We have them record the data using Mathematica, plot the results, and produce a nonlinear fit to the theory curve. The exercise teaches self-sufficiency in data handling and analysis skills, as well as providing some experience using a popular software tool.

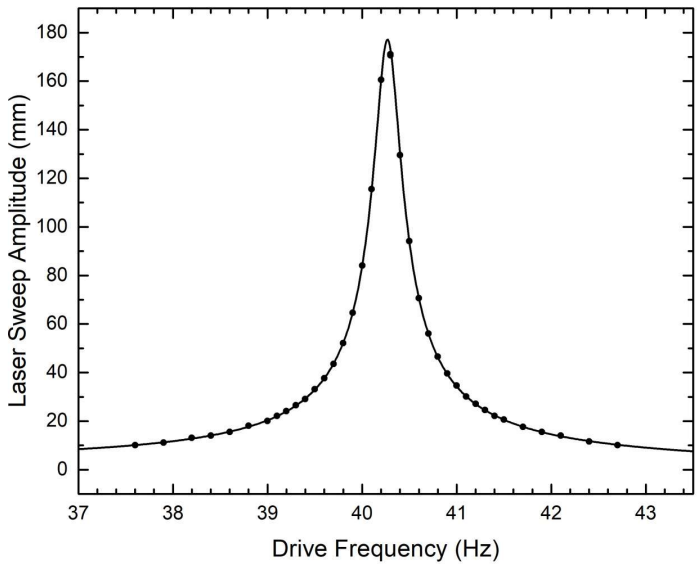

Figure 9. This plot shows the length of the laser streak on a plastic ruler (see Figure 3) as a function of the frequency of a constant-amplitude sinusoidal voltage signal applied to the Drive Coil, with the ruler placed about one meter from the MMHO. A measurement accuracy of about 1mm can be obtained visually, and the resulting data points are nicely described by the fit line using Equation (9). Here the fit yielded $f_0 \approx 40.267 \pm 0.003$ and $Q \approx 280 \pm 10$.

### From the photodiode signal

If one wants to take this to the next level, a data-acquisition system can be used to record the PD signal as a function of the drive frequency to measure the MMHO response. We used a Rohde&Schwarz MXO4 oscilloscope with a Frequency Response Analysis (FRA) feature for this task, which is a great tool that is probably too expensive for most teaching-lab budgets. However, oscilloscopes with FRA capabilities are coming down rapidly in price, so this may be a viable option for many teaching labs in the near future. Alternatively, computer-based systems like LabView can be used to record the PD signal and calculate its amplitude and phase.

To produce the data shown in Figure 10, we set up the MMHO with the eddy-current damper in place and set up the FRA system to produce a series of measurements with a settling time of 10 seconds after each frequency change. It took about 20 minutes to collect the data in each plot in Figure 10. Overall, the MMHO clearly performed well over a large dynamic range, but we did observe a couple areas in which the data did not fit SHO theory:

1) At high frequencies, the data points do not match the theory line because of a broad mechanical resonance in the MMHO tower at about 180 Hz. This non-ideal SHO behavior could probably be reduced by building a stiffer tower… a project left for another day.

2) The phase data exhibited a constant offset compared with theory, and we added a constant phase offset of 7.6 degrees to the data points in Figure 10 to compensate for this effect. The source of the offset was traced to a variety of electronic phase lags in MMHO system. For example, the Drive Coil inductance produced a phase lag in the coil current relative to the applied drive voltage. In addition, the PD signal electronics introduced another small phase shift in the PD signal relative to the optical input. These phase shifts could be reduced with better electronics design... again a project left for another day.

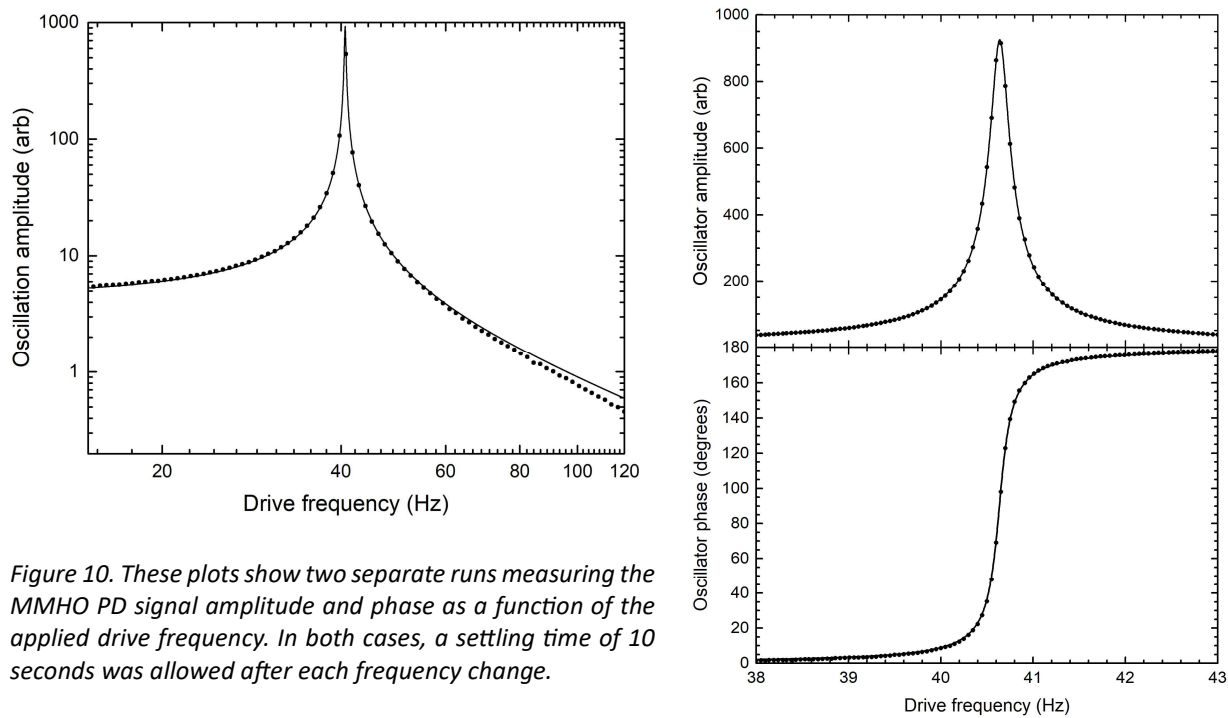

*Figure 10. These plots show two separate runs measuring the MMHO PD signal amplitude and phase as a function of the applied drive frequency. In both cases, a settling time of 10 seconds was allowed after each frequency change.*

All things considered, the MMHO response is remarkably well described by SHO theory, but there is still room for improvement in the design.

### Budget FRA

Frequency-response analysis can even be demonstrated with a basic oscilloscope, following the discussion above with ringdown measurements. The basic idea is to sweep the drive frequency slowly over time (using the Sweep feature in most modern signal generators) and then observe the PD signal using the oscilloscope in Roll mode. For the data shown in Figure 11, we used a Rigol DHO 804 oscilloscope, which has a 12bit input resolution and a 25M memory depth at a current list price of just $329. Recording 100,000 data points is sufficient to collect several points per oscillation cycle over eight minutes, and handling such a data set is manageable using a basic PC.

One disadvantage of this budget FRA method is that it is difficult to produce sweeps that are slow enough to give the oscillator time to relax to its steady-state amplitude. Figure 12 illustrates how the peak becomes quite distorted at high sweep speeds. (The algorithm used to smooth the PD data produced a negligible shift in these data.)

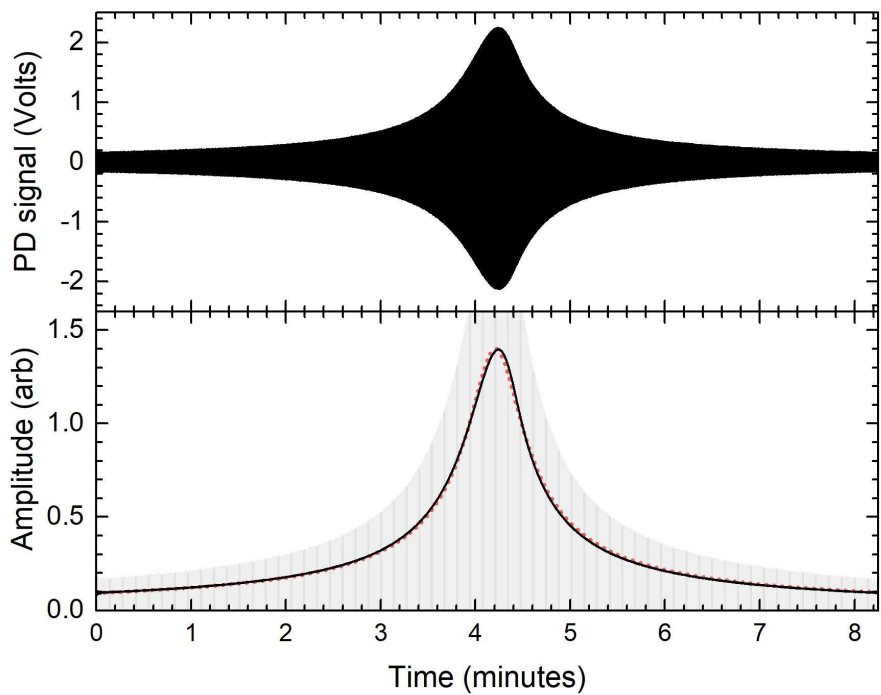

*Figure 11. (Top) This plot shows the raw oscilloscope data, recording the PD signal as a function of time with the oscilloscope in Roll Mode. Individual oscillation cycles are resolved in the raw data, but not in this plot. The frequency was swept linearly over a range of 2.5 Hz during this time at a rate of 5 mHz/sec, showing a peak PD signal amplitude at $f_0$. (Bottom) The black line in this graph shows a smoothed version of the absolute value of the amplitude signal, and the red dotted curve shows a best fit to theory. The measured peak is shifted slightly because the oscillator amplitude did not quite come to its steady-state value during the sweep.*

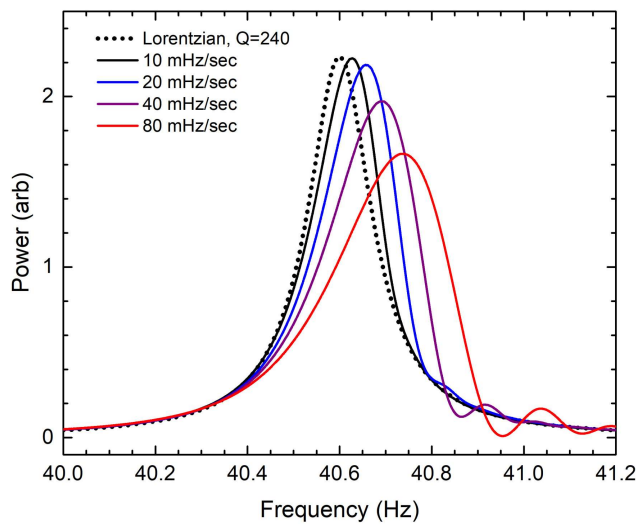

*Figure 12. Here we swept the drive frequency at different rates and observed how the MMHO response peak shifted with increasing sweep speed.*

**Transient behaviors**

The oscilloscope's Roll Mode can also be used to observe a variety of thought-provoking transient effects in the MMHO behavior. If the oscillator starts out at rest and a drive signal is abruptly turned on with a frequency is equal to $f_0$, then the amplitude will relax exponentially to its steady-state value, exhibiting a behavior that is analogous to that seen in Figure 6. At any other drive frequency, the oscillator exhibits a complex transient behavior that includes a "beat-note" response at a frequency equal to $f - f_0$, as illustrated in Figure 13.

Modeling these transient behaviors is nontrivial, but it is a good example of what one can do with a mathematically capable AI system. SHO physics is sufficiently documented that the AI likely knows how to solve the equation of motion, making it possible to simply ask what problem you would like to solve and plot. The current low-price AI tools (at the time of this writing) may not get these calculations right, but they are improving quickly. The transient SHO problem is complex enough to be interesting, yet easy enough to solve using basic AI capabilities. Having laboratory examples of complex transient waveforms available on demand provides an interesting exercise in using AI for problem solving.

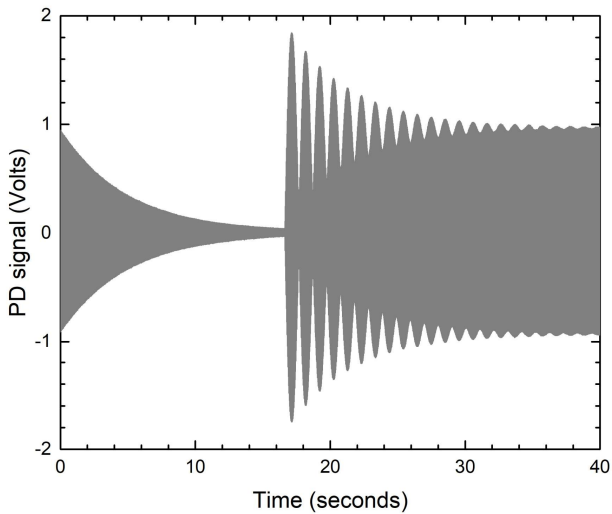

*Figure 13. In this oscilloscope trace, we first let the oscillator relax down to near-zero amplitude and then applied a drive signal at 1Hz above $f_0$. The oscillator responded with a 1Hz modulation of the amplitude that eventually settled to its steady-state value. This Roll-Mode technique can be used to explore a variety of transient oscillator behaviors.*

**With minimal damping**

The capabilities of the MMHO instrument become especially evident at high Q values, and Figure 14 illustrates this with a measurement of the oscillator frequency response. We again used the Rohde&Schwarz MXO4 oscilloscope to collect these data, and at 10 seconds/point it took over eight hours to collect the 3000 data points in the main plot. While this exercise is clearly beyond what is normally seen in physics teaching labs, it is quite fascinating to see such precise data spanning five orders of magnitude of dynamic range coming from such a simple instrument.

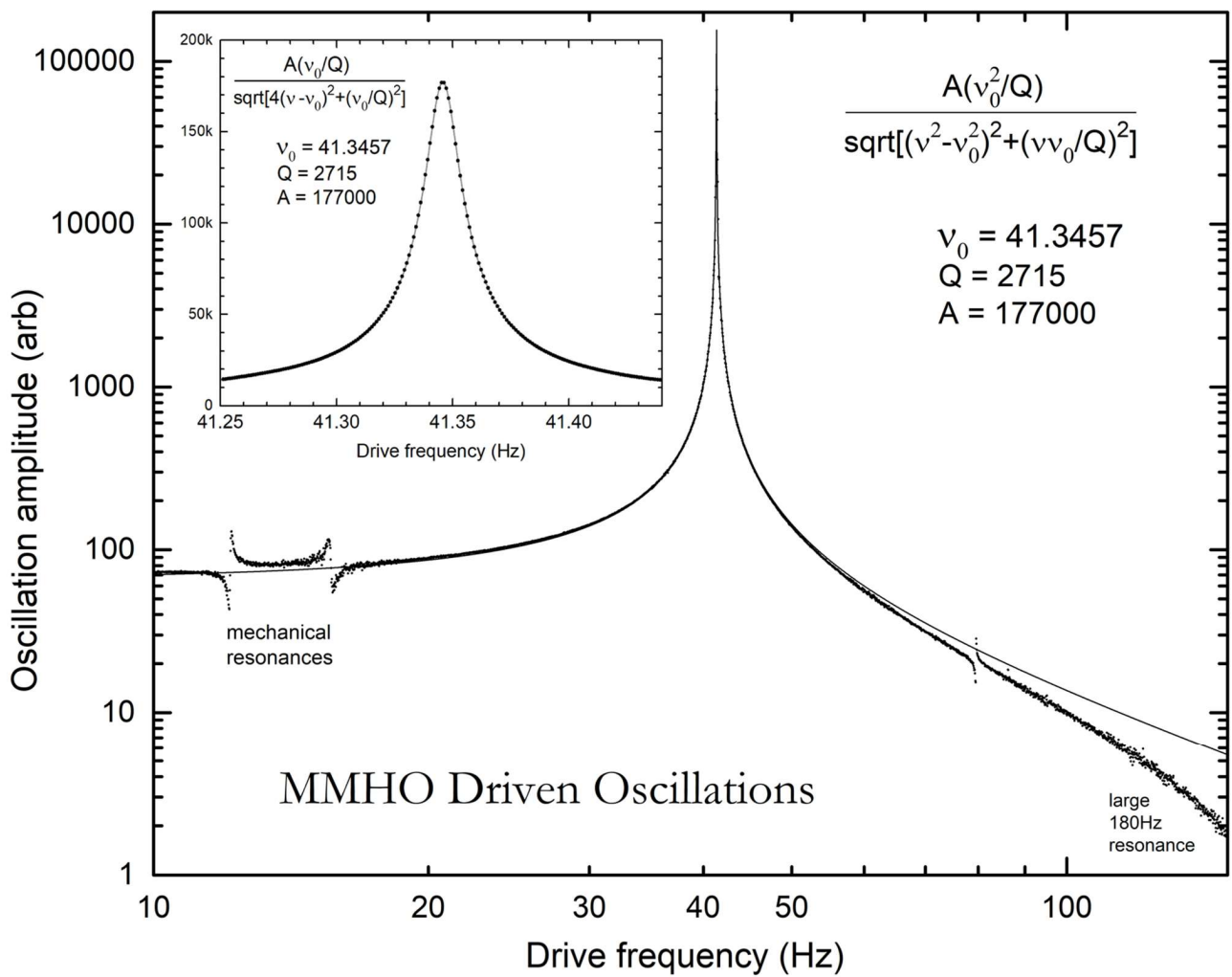

*Figure 14. This plot shows the MMHO response with the eddy-current damping removed, yielding a sharp resonant peak and measurements spanning a remarkable five orders of magnitude in amplitude. The measured response reproduces SHO theory nearly exactly around resonance, with unwanted mechanical resonances appearing only far from resonance and at low amplitudes. The insert shows a separate run to obtain higher resolution near the resonance peak.*

## Clock drive

We also included a "self-excitation" mode in the MMHO that uses positive feedback of the PD signal to drive the oscillator. We call this "Clock Drive" because this is essentially how all clocks work. In a nutshell, the MMHO electronics turns the PD sinusoid into a square wave, takes the time derivative of this signal to produce a series of voltage spikes, and one then sends these spikes to the MMHO Drive Coil. Figure 15 illustrates these signals, and the electronic circuit that produces the Clock Drive signal is provided in Appendix 1. Once the Clock Drive is engaged, the MMHO oscillator soon reaches a steady-state amplitude that can be adjusted using a gain knob that amplifies the size of the clock-drive signal. Note that the drive spikes occur when the test mass position angle is near $\theta = 0$, with the oscillator evolving freely between these spikes. Because the drive spikes are derived from the PD signal, the clock drive produces oscillations at the natural resonant frequency $f = f_0$.

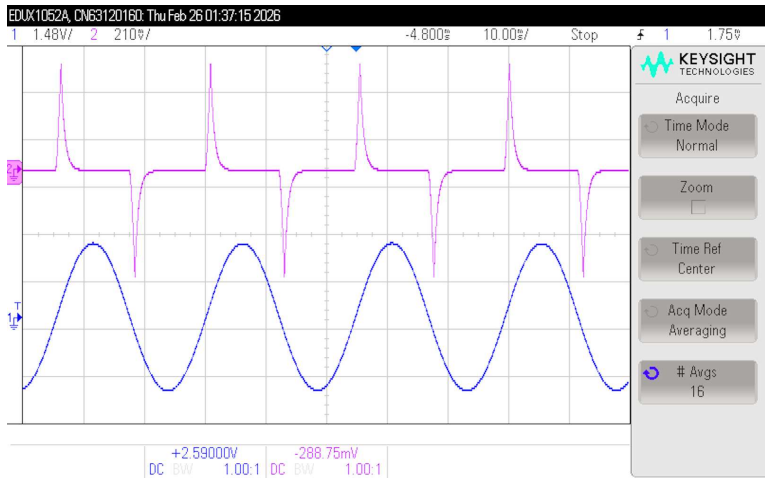

*Figure 15. This oscilloscope screenshot shows the Clock Drive signal (top trace) that is derived from the PD signal (bottom trace). The electronic circuit that produced the Clock Drive signal is shown in Appendix 1.*

## Measuring the resonance frequency

In keeping with a common theme in this paper, there are several methods one can use to measure the natural resonant frequency $f_0$ of the MMHO. One easy method is to simply drive the oscillator using an external function generator and adjust the drive frequency to maximize the length of the laser streak. This method is quick and visual, and it is possible to measure $f_0$ to an accuracy of ±50 mHz if one is careful.

Another method for measuring $f_0$ using the laser streak is to measure the oscillation amplitude as a function of frequency and then fit the resulting data, as illustrated in Figure 9. This method is somewhat tedious, but accuracies of ±10 mHz can be obtained.

With the Clock Drive engaged, another way to measure $f_0$ is a direct frequency measurement of the PD signal using the Measure feature on an oscilloscope. This method typically yields accuracies of ±100 mHz using a low-end oscilloscope, but better accuracy can be obtained using more expensive models. Averaging many measurements can yield ±5 mHz or better, but the ability to collect measurement statistics is not always available on low-end 'scopes.

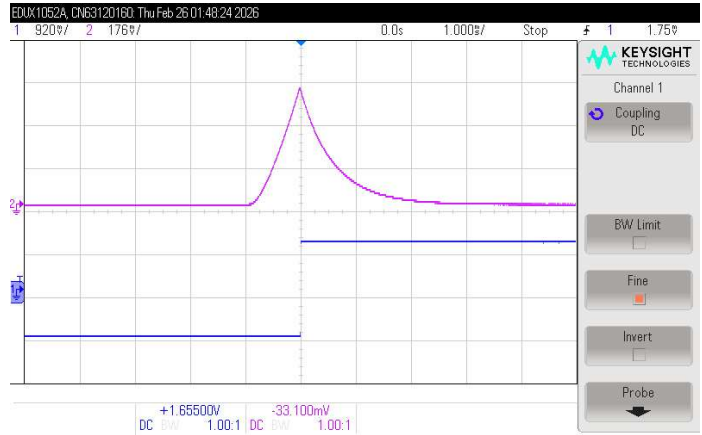

*Figure 16. This oscilloscope screenshot shows a single spike in the Clock Drive signal (top trace) together with the rising edge of a square-wave signal (produced by a signal generator). If the two signals do not have the same frequency, then one moves across the screen relative to the other. When both signals are stationary, the signal-generator frequency equals $f_0$.*

Strobing the laser using a signal generator while the MMHO is in clock mode provides another fun and visual technique for measuring $f_0$. One simply adjusts the strobe frequency until the short streak of the strobed laser is stationary in time, and measurement accuracies of $\pm 10$ mHz are easily obtained.

Yet another method for measuring $f_0$ is to view the Clock Drive signal on the oscilloscope along with a square wave signal provided by a signal generator. With the 'scope triggered on the square wave, one then lines up the two signals so the clock-drive signal does not move relative to the square wave, as illustrated in Figure 16. By zooming in on a single clock-drive pulse, this method can yield accuracies of $\pm 1$ mHz with only moderate effort.

If you want to provide the full professional-scientist experience, then the best method for measuring $f_0$ is by simply using a frequency counter to measure the PD signal frequency with the MMHO in clock-drive mode. The Keysight 53220A Universal Counter/Timer is a good choice, although cheaper frequency counters (for example the VQP 10HZ-2.7GHZ High Resolution Frequency Counter, which has a price of ~$100) will still deliver <10ppm accuracy. The Keysight 53220A offers data-logging capability as well, and Figure 17 shows a long timeseries of observations of the MMHO in clock drive.

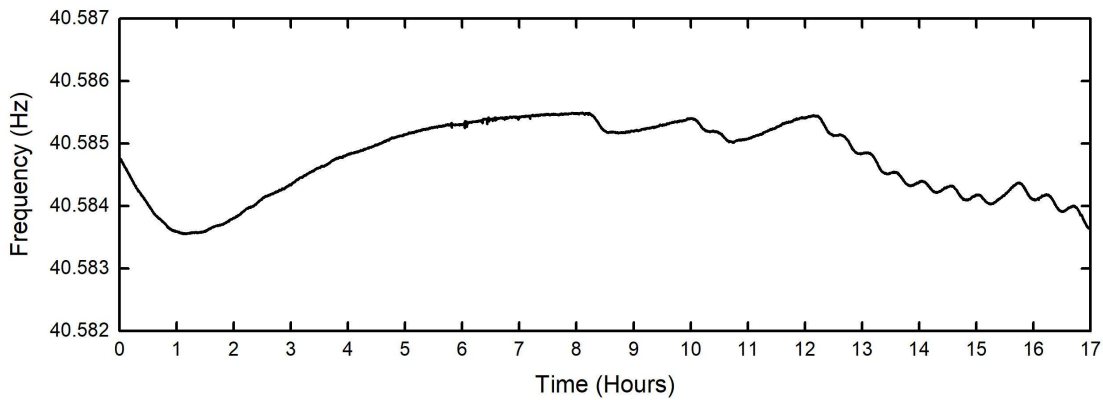

*Figure 17. These data show the measured $f_0$ as a function of time, demonstrating ±1 mHz stability over many hours. The frequency drift mainly arises from changes in the MMHO spring constant with temperature. Used as a timepiece, the MMHO would be accurate to a few seconds per day – poor by modern clock standards, but not bad for a teaching-lab mechanical oscillator.*

## Frequency shifts

Because the oscillation frequency is quite stable and can be easily measured with high accuracy, we found it interesting to measure how $f_0$ varies with other MMHO parameters. Figure 18 shows some data, and the challenge then becomes developing plausible theoretical explanations for all the features seen in these observations.

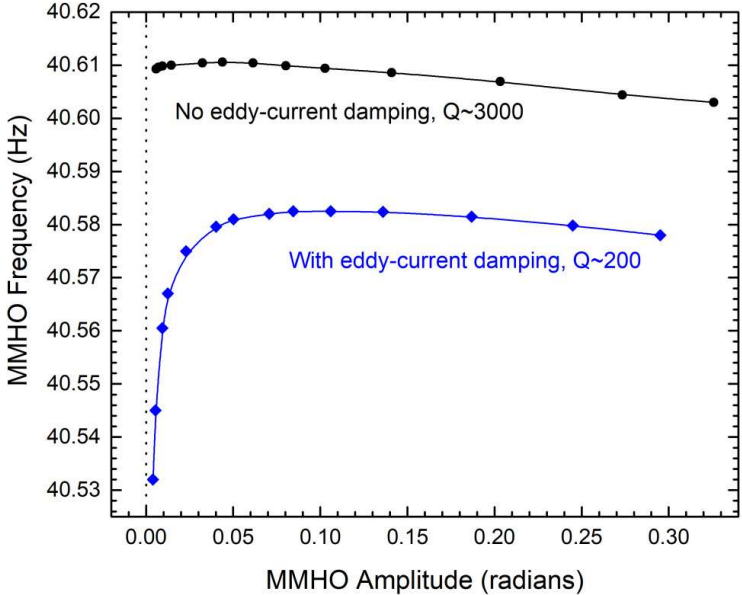

*Figure 18. These data points show measurements of the MMHO oscillation frequency $f_0$ in clock-drive mode as a function of the oscillator amplitude $\theta_0$, both with and without eddy-current damping. The curves are spline fits to the data points, meant to guide the eye.*

For example, in a perfect SHO we expect that $f_0$ should be independent of the oscillation amplitude [2026Riz], while Figure 18 reveals (among other things) that $f_0$ decreases slowly with increasing amplitude at high amplitudes. Frictional heating in the MMHO support wires may explain this trend, as higher amplitudes produce more wire heating and this may affect the spring constant $\kappa$. Looking into this possibility, a bit of research reveals that $\kappa$ for a steel spring is proportional to the Young's modulus $E$ of the material, and $(1/E)dE/dT \approx 2 \times 10^{-4}/C$ is typical for steel. Putting in some numbers, the data in Figure 18 could be explained by a temperature increase of roughly one degree at the highest amplitudes. We cannot definitively confirm this hypothesis, however, and other effects may be more important than wire heating.

### Phase shifts in eddy-current damping

The change in $f_0$ with eddy-current damping in Figure 18 presents another interesting observation. Equation (7) predicts a $\Delta f_0 \approx 0.5$ mHz when the damping is increased to produce $Q \approx 200$, while the data revealed $\Delta f_0 \approx 30$ mHz. This sizable discrepancy might be explained by phase shifts in our physical model of eddy-current damping. The damping parameter in Equation (2) is appropriate for normal frictional damping, which acts essentially instantaneously because atomic motions are normally much faster than motions of a macroscopic test mass like that in the MMHO. But eddy-currents are macroscopic electrical currents generated in the copper damper, and electrical currents often exhibit phase lags relative to the EMFs that cause them.

To include this possibility, let $\theta(t) = \sin(\omega_0 t)$ and modify Equation (2) to

$$I\ddot{\theta} \approx -\kappa\sin(\omega_0 t) - \gamma\omega_0\cos(\omega_0 t - \varphi) \tag{11}$$

where $\varphi$ postulates a phase lag in the eddy-current damping forces. Expanding this expression for $\varphi \ll 1$ yields

$$I\ddot{\theta} \approx -(\kappa - \gamma\omega_0\varphi)\sin(\omega_0 t) - \gamma\omega_0\cos(\omega_0 t) \tag{12}$$

and we see that a phase lag produces (to first order) a small reduction in the effective spring constant of the oscillator. In this picture, the oscillation frequency becomes

$$f_0 \to f_0\left(1 - \frac{\gamma\omega_0\varphi}{2\kappa}\right) \tag{13}$$

and explaining the observations requires a phase lag of $\varphi \approx 20$ degrees. Once again, this phenomenon deserves additional study, but we postpone that for another time.

We have not found a ready explanation for the remarkable drop-off in $f_0$ at low amplitudes when eddy-current damping is present, as seen in Figure 18. A detailed physical model of eddy-current damping would be helpful here, but the induced currents involved are far from simple.

Overall, these observations of small changes in $f_0$ with damping and amplitude are probably unsuitable for most laboratory teaching applications, as they distract from the main event of SHO physics. But these musing do demonstrate that the MMHO need not be limited to introductory laboratory teaching. When you look closely enough at a precision instrument, interesting new physics often comes into play.

## Characterizing the MMHO

Now that we have seen what the MMHO can do, it is a useful exercise to examine the various physical parameters that characterize its operation. For the MMHO, we can begin by noting that a measurement of the oscillation frequency gives us

$$\frac{\kappa}{I} = 4\pi^2 f_0^2 \tag{14}$$

while a ringdown measurement gives the ratio

$$\frac{\gamma}{I} = T_{ringdown}^{-1} \tag{15}$$

Combining these expressions yields another ratio

$$\frac{\kappa}{\gamma} = 4\pi^2 f_0^2 T_{ringdown} \tag{16}$$

To isolate the individual parameters in Equation (2), we can estimate $I$ from the moment of inertia of a cylinder in this orientation

$$I = \frac{1}{12}M(3R^2 + L^2) \tag{17}$$

where $R = 6.35$ mm, $L = 19.25$ mm, and $M = 18$ grams for the MMHO test magnet, giving $I_{magnet} = 7.37\times 10^{-7}$ kg-m$^2$ just for the test-mass magnet. The rest of the test-mass assembly adds about an additional five percent, for a total $I_{test-mass} \approx (7.7 \pm 0.2) \times 10^{-7}$ kg-m$^2$.

On the other side of Equation (2), we would also like to know the value of the magnetic moment $\mu$ of the test mass magnet, and measuring this requires the application of a known magnetic field. The geometry of the Bias Coils (see Figure 1) is particularly simple and therefore best suited for this task, and a straightforward analysis of the specific MMHO coils give $B_{bias} = \beta I_{bias}$ with $\beta \approx 5.1 \pm 0.2$ mT/Amp at the position of the test-mass magnet.

The Bias Coils produce a magnetic field along the magnetic axis of the test mass (when it is at rest), thus exerting a torque $\tau = \mu \times B_{bias} = \mu B_{bias} \sin(\theta) \approx \mu B_{bias} \theta$, again using the small angle approximation. When the bias coils are energized, the spring constant of the oscillator thus becomes $\kappa = \kappa_0 + \mu B_{bias}$, where $\kappa_0$ is the spring constant from the wires alone, and the MMHO oscillation frequency becomes

$$4\pi^2 f_0^2 I = \kappa_0 + \mu B_{bias} \quad (18)$$

Figure 19 shows experimental MMHO data, and putting in the numbers gives $\mu \approx 2.2 \pm 0.15$ Amp-m$^2$. (Note that 1 Amp-m$^2$ is the same as 1 Joule/Tesla, and both these SI units are commonly used.) A bit of research suggests that a typical rare-earth magnet has a magnetic moment per unit volume of about $10^{-6}$ Amp/m, giving an expected $\mu \approx 2.4$ Amp-m$^2$, so our measurement is in reasonable agreement with expectations.

As a side note, the data point in Figure 19 are remarkably precise because both the applied current and the MMHO oscillation frequency can both be measured to an absolute accuracy of about one part per thousand. Subtracting the linear fit in Figure 19 yields residuals that are not random, but clearly show nonlinear effects (one obvious source being deviations from the small-angle approximation used in deriving Equation (18)). Exploring these nonlinearities in detail is yet another exercise we leave for another day.

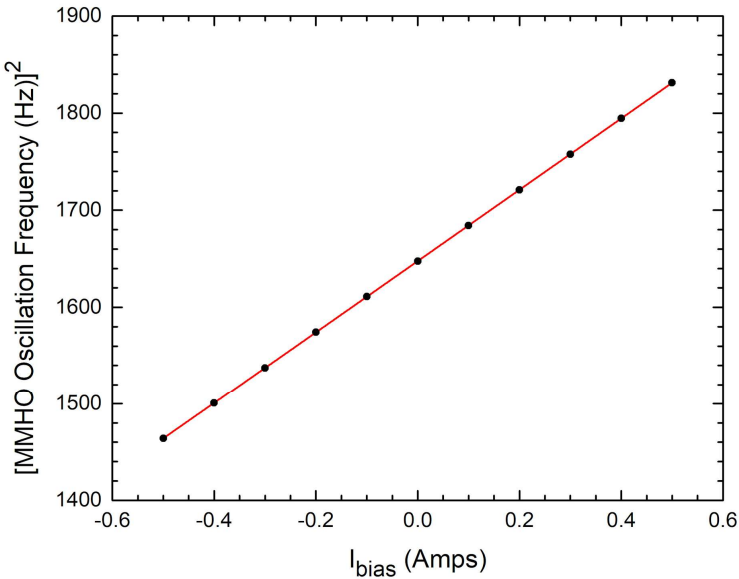

Figure 19. Measurements of the MMHO oscillation frequency (squared) as a function of the current in the Bias Coils, giving a slope of 367 $Hz^2$/A. The data points here are remarkably well defined, as both the MMHO oscillation frequency and the applied current can be measured to an absolute accuracy of about one part per thousand.

## Parametric drive

The Bias Coils also provide an easy way to observe *parametric excitation* of the MMHO oscillator, which is a fascinating phenomenon produced by applying a sinusoidal modulation of the spring constant $\kappa$ at a frequency equal to $2f_0$. A full mathematical derivation of the theory can be provided by your favorite AI tool, as the subject is well documented online. But it is pedagogically useful to consider an approximate physical derivation here.

As described above, a static magnetic field from the Bias Coils produces a torque on the MMHO test mass equal to $\tau \approx \mu B_{bias} \theta$ in the small-angle approximation. Now consider what happens when the Bias Coils are driven at $2f_0$ with $B_{bias}(t) = B_1 \cos(2\omega_0 t)$ while the MMHO happens to be oscillating in synchrony at its normal oscillation frequency with $\theta(t) = \theta_1 \cos(\omega_0 t)$. At $t = 0$, we see that $B_{bias}$ is fully on while $\theta$ is at its maximum extension. As a result, the test mass experiences a magnetic torque pulling it back home at that time. And the same thing happens at $t = \pi/\omega_0$. Thus, for this totally contrived initial condition, the test mass receives two "kicks" during each oscillation cycle from the Bias Coils, and both kicks pump energy into the oscillator. The magnitude of each kick is $\tau_{mag} \approx \mu B_1 \theta_1$.

While these magnetic kicks are driving the oscillator amplitude higher, normal frictional forces tend to drive the amplitude lower. The frictional torques are highest when the oscillator is moving the fastest, and Equation (2) tells us that the maximal frictional torque is given by $\tau_{drag} \approx \gamma \dot\theta \approx \gamma \omega_0 \theta_1$. Once again, we see that the oscillator receives two damping "kicks" during each oscillation cycle. Importantly, both $\tau_{mag}$ and $\tau_{drag}$ are proportional to $\theta_1$.

Putting the pieces together, we see that the oscillator is continuously receiving kicks from the Bias Coils that tend to speed it up and additional kicks from friction that tend to slow it down. In a rough approximation, if $\tau_{drag} > \tau_{mag}$, then friction wins and the oscillator amplitude decays with time. But if the opposite inequality is true, then the Bias Coils win and the oscillator amplitude increases with time. The net change in $\theta_1$ with time can be written

$$\frac{d\theta_1}{dt} \approx C(\mu B_1 - \gamma \omega_0)\theta_1 \qquad (19)$$

where $C$ is a constant that depends on the impact of each kick. And this differential equation describes either exponential decay or exponential growth, depending on the sign of the inequality.

And now you can see that our initial starting point of synchronous motion was not important. The synchronous mode will grow exponentially if $B_1$ is high enough, while other oscillation phases will not. If we start with the oscillator at rest, then it will remain at rest below some threshold $B_1$. Above this threshold, the oscillator amplitude in the synchronous mode will grow exponentially, and the speed of the exponential rise will depend on how high the drive is above threshold.

A full derivation of this parametric excitation is straightforward but somewhat tedious, starting with Equation (2) and doing careful averages of all the relevant forces. The result is similar to Equation (19), including a calculated value for $C$.

The MMHO provides a quick and easy demonstration of this phenomenon, and Figure 20 shows the predicted exponential increase in oscillator amplitude as a function of time. The threshold in this case was about 90 mVrms, and higher applied fields indeed yielded faster exponential growth. Although the PD signal saturates above about 3 volts (see Figure 4), the oscillator amplitude continued to grow exponentially until we turned the Bias Coils off, worried that such high amplitudes might damage the instrument. It is an impressive experience to watch the oscillator amplitude rise ever so slowly at first, then faster and faster at later times.

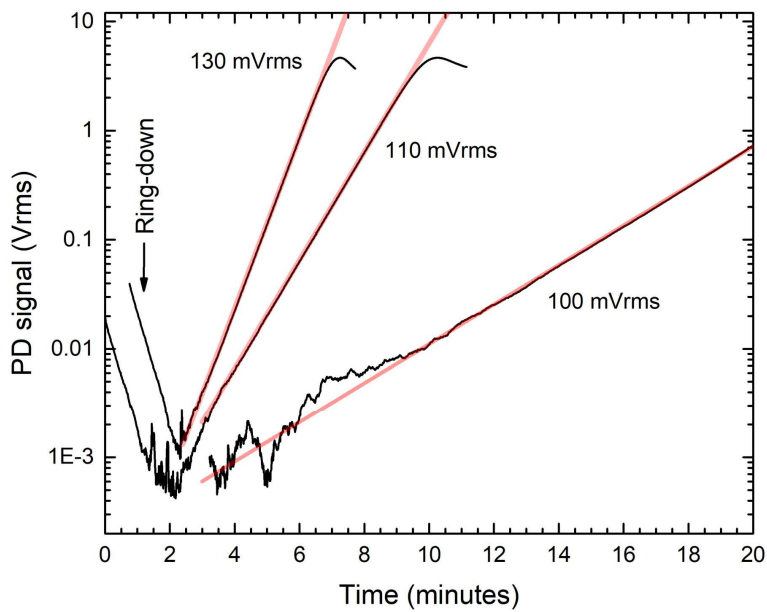

*Figure 20. These data (black lines) show three separate runs measuring the PD signal amplitude as a function of time. Each trace begins with a ringdown period, during which the amplitude decayed exponentially. When the signal was around 1mV, the Bias Coils were driven at a frequency of $2f_0$ with the voltage amplitudes shown. In each case the PD signal increased exponentially with time (until nonlinear effects in the PD signal dominated above 3Vrms). With a drive below 90mVrms, no exponential increase was observed. The measured exponential time constants (red lines) were 2.40, 0.875, and 0.555 minutes.*

## Summary

One strength of the MMHO instrument is that it can perform a wide variety of different experiments exploring the physics of simple harmonic motion. The apparatus also provides a fun platform for teaching students about optics, electronics, data acquisition, and general tools and techniques one encounters in experimental physics. Most of the experiments described above are aimed at the intro-lab level, although some of the physics is more subtle. Our mission in this paper is to describe what an instrument like the MMHO is capable of, leaving it to lab instructors to decide what materials are best suited for their students, and whether building their own MMHO is something they want to consider.

## Acknowledgements

This work was supported in part by a generous donation from Beatrice and Sai-Wai Fu to the Physics Teaching Labs at Caltech. We also acknowledge support from Drs. Richard Karp and Vineer Bhansali, together with Caltech's long-standing support of outstanding laboratory instruction across many STEM fields.

*Contact:* For corrections, comments, or just to compare notes, please contact Kenneth G. Libbrecht, *kgl@caltech.edu*, or mail to: Mail-stop 264-33 Caltech, Pasadena, CA 91125.

# Appendix 1 – Hardware details

The main parts of the MMHO construction include an acrylic tower bolted to an electronics chassis, as shown in Figure 1. The chassis was a heavy steel box designed for this purpose using *Protocase* design software, including an array of mounting holes on the top surface. The *Protocase* box has a footprint of 26x28 cm and was made from thick steel to provide a heavy and stable base. Being far from the test mass, this steel provided negligible eddy-current damping.

The main tower is made from four layers of 6mm laser-cut acrylic, and (in hindsight) this could have been made thicker and sturdier to avoid unwanted mechanical resonances. We chose plastic construction here in part to avoid unnecessary eddy-current damping. The side towers holding the laser and photodiode assembly were made from 12.7mm acrylic square tubes welded into laser-cut acrylic bases.

The test-mass magnet is described in the main text above, and it is glued into another laser-cut acrylic plate. The test-mass mirrors are 12.7mm in diameter and 1.6mm thick with front-surface aluminum coatings. A high degree of flatness was not needed with these mirrors. The tension wires were made from #2 gauge music wire, each wire being about 8.5cm in length with the ends silver-soldered into the heads of stainless-steel bolts. The choices of high-carbon steel and silver-soldered construction were made to reduce frictional losses in the wires, yielding Q values over 3000 in some cases (when no eddy current damping is applied). We tested one oscillator by letting it run for over a month with continuous high-amplitude oscillations, and we saw no signs of metal fatigue in the wires at the end of this test.

The mounting of the wire ends in the acrylic towers turned out to be one of the trickier aspects of the MMHO. If there is too much tension in the wires, the tower exhibits some shaking and the Q is reduced. If there is too little tension, the test mass becomes a bit "floppy" in its operation. Clamping the wire ends in the tower to obtain good results is quite easy, but the tension often relaxes with time and usage. We have not solved this problem fully, but we settled on a reasonable compromise. For low-Q use, we developed a "self-tensioning" system by hanging a weight on the lower end of the assembly. And for high-Q use we simply clamp the ends tightly (which holds for weeks or months, but not forever).

The Drive Coil was made from 300 turns of #30 gauge magnet wire, while each Bias Coil was made from 200 turns of #28 gauge magnet wire. The MMHO electronics schematic shown in Figure 21 shows the circuits inside the main chassis. Another design failing (in hindsight) that we would correct in another design iteration is non-negligible phase lags that arise in the Drive Coil and PD amplifier circuits. These lags could be greatly reduced with better circuit design (for example using an op-amp based high-speed current driver for the Drive Coil circuit), allowing the driven-oscillator phase measurements to yield better agreement with theory.

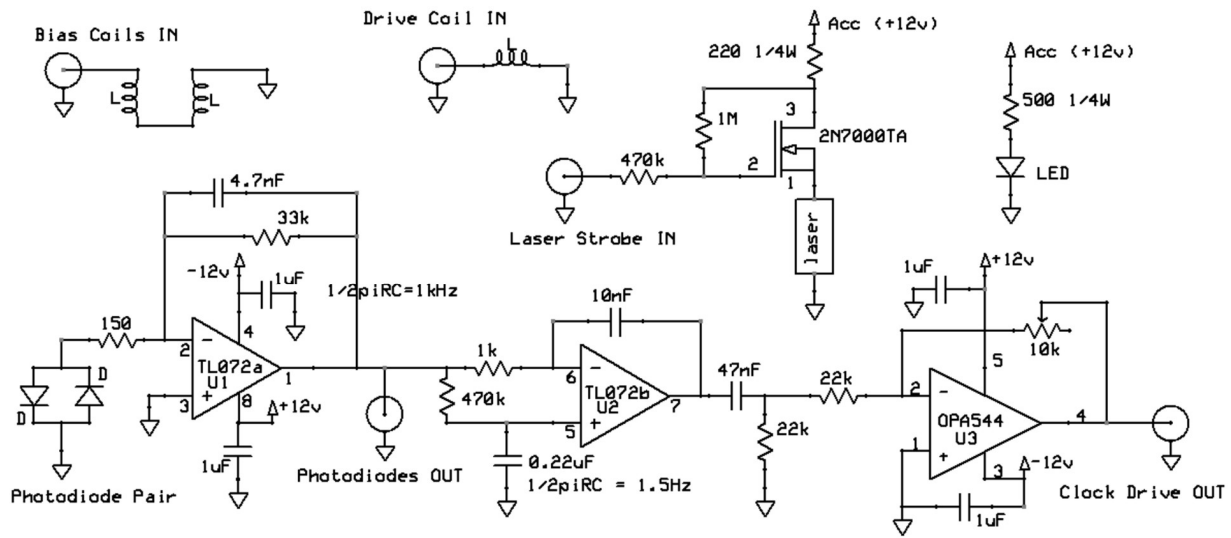

*Figure 21. The MMHO electronics schematic, showing the different circuits and BNC outputs on the front chassis.*